\documentstyle[epsbox,12pt]{article}
\newcommand{\la}{\langle}
\newcommand{\ra}{\rangle}
\begin{document}
\title{KNO scaling function of modified negative binomial distribution}
\author{Noriaki Nakajima, Minoru Biyajima, 
         and Naomichi Suzuki$^{1}$ \\
 ${}^{}$Department of Physics, Shinshu University, Matsumoto 390, 
Japan \\
 ${}^{1}$Matsusho Gakuen Junior College, Matsumoto 390-12, Japan }  
\date{}
\maketitle
\begin{abstract}
We investigate the KNO scaling function of the modified 
negative binomial distribution (MNBD), because this MNBD can explain
the oscillating behaviors of the cumulant moment observed
 in $e^+e^-$ annihilations and in hadronic collisions.
By using a straightforward method and the Poisson transform we 
derive the KNO scaling function from the MNBD.
 The KNO form of experimental data in $e^{+}e^{-}$ collisions and 
 hadronic collisions are analyzed by the KNO scaling function of the MNBD 
 and that of the negative binomial distribution (NBD). 
The KNO scaling function of the MNBD describes the 
data as well as that of the NBD.
\end{abstract}

\section{Introduction}

Recently it has been found that cumulant moments of the multiplicity 
distributions both in $e^+e^-$ annihilations and hadronic collisions
show prominent oscillatory behaviors when plotted as a function of their
order $q$ \cite{DREMIN}. In \cite{DREMIN} this behavior was attributed
to the QCD-type of branching processes apparently taking place in
those reactions. However, in \cite{NBS,NBSh} we have shown that
the same behavior of the moments emerges essentially from the modified 
negative binomial distribution (MNBD) (which actually describes
the data much better than the negative binomial distribution (NBD)
\cite{MNB}). This distribution can be derived from the pure birth 
process with the initial condition given by the binomial distribution
\cite{BSW}\footnote{It is interesting to mention here that it comes
also from the concept of purely bosonic sources as presented recently in
\cite{BSWW}.}. \\

In this paper we shall derive the KNO scaling function of the MNBD
both by the straightforward method (i.e., proceeding to the limit 
of large multiplicities $n$ and large average multiplicities 
$\langle n\rangle$ while keeping the scaling variable
$z=n / \langle n\rangle $ finite and fixed) and by using 
the Poisson transform.
Using this KNO scaling function we shall analyze the observed multiplicity
distributions in $e^{+}e^{-}$ annihilations and in hadronic
collisions.

\section{KNO scaling function}

Let us remind that the MNBD is given by
the following function \cite{NBS,NBSh,MNB,BSW} 
\begin{eqnarray}
  P(0) &=& \left[ \frac{1+r_{1}}{1+r_{2}} \right]^N ; \nonumber
            \\  
  P(n) &=& \frac{1}{N!}\left(\frac{r_{1}}{r_{2}}\right)^N
           \left(\frac{r_2}{1+r_2}\right)^n
           \sum_{j=1}^{N} {}_{N}C_{j} \frac{\Gamma (n+j)}{\Gamma (j)}
           \left( \frac{r_2-r_1}{r_1} \right)^j
           \frac{1}{(1+r_2)^j} , \label{eq:2}
\end{eqnarray}
where $N$ (the number of excited hadrons) is an integer, $r_{1}$ is real and $r_{2}>0$
\begin{eqnarray}
  r_{1} &=& \frac{1}{2} \left(C_{2}-1-\frac{1}{N} \right)\langle n \rangle-
           \frac{1}{2} ,
           \nonumber \\ 
  r_{2} &=& \frac{1}{2} \left(C_{2}-1+\frac{1}{N} \right)\langle n \rangle-
           \frac{1}{2}.  \label{eq:3}
\end{eqnarray}
Moreover we have the generating function of the MNBD
\begin{equation}
	\Pi(u) = \sum_{n=0}^{}P(n)u^{n}
	       = [1-r_1(u-1)]^{N}[1-r_2(u-1)]^{-N}.
	\label{eq:0}
\end{equation}

On the other hand, the NBD has the following form
$$P(n)=\frac{\Gamma (n+k)}{\Gamma (n+1) 
\Gamma (k)} \left( \frac{k}{\langle n \rangle} \right)^{k} \left(1+ 
\frac{k}{\langle n \rangle} \right)^{-(n+k)},$$
where $k>0$.
Its corresponding KNO scaling function is the gamma distribution
\begin{equation}
   \Psi(z)=\frac{k^k}{\Gamma (k)}e^{-kz}z^{k-1}.\label{eq:4}
\end{equation}
\subsection{The straightforward method}

Traditionally the KNO scaling function is derived from the multiplicity
distribution $P(n)$ multiplied by the corresponding mean multiplicity 
$\la n\ra$ by going to the large multiplicity $n$ and large mean 
multiplicity $\la n\ra$ limit while keeping their ratio, 
$ z = \lim_{n,\langle n \rangle\rightarrow 
\infty}n/\langle n \rangle$ fixed. In our case, starting 
from Eq. (\ref{eq:2}) we arrive at the following function 
\begin{eqnarray}
        \Psi(z)  & \equiv & \lim_{n,\langle n \rangle\rightarrow 
\infty}\langle n \rangle P(n)
                       \nonumber \\
                 & = & \left(\frac{r'_1}{r'_2}\right)^{N}
             e^{-\frac{\langle n \rangle}{r'_2}z}
             \sum_{j=1}^{N}{}_NC_{j} \frac{1}{\Gamma(j)}
             \left(\frac{r'_2-r'_1}{r'_1}\right)^{j}
              \left(\frac{\langle n \rangle}{r'_2} \right)^{j}z^{j-1}.
        \label{eq:5}
\end{eqnarray}
The parameters $r'_1$ and $r'_2$ in Eq. (\ref{eq:5}) are given by
\begin{eqnarray}
  r'_{1} &=& 
    \frac{1}{2} \left(C_{2}-1-\frac{1}{N} \right)\langle n \rangle
           \nonumber \\ 
  r'_{2} &=& 
    \frac{1}{2} \left(C_{2}-1+\frac{1}{N} \right)\langle n \rangle,
 \label{eq:06}
\end{eqnarray}
which are slightly different from 
$r_1$, $r_2$ given by Eq. (\ref{eq:2}), because $\la n \ra \gg 1$.
It should be noticed that the normalization of Eq. (\ref{eq:5}) 
differs now from the unity,
\begin{equation}
        \int_{0}^{\infty}\Psi(z)dz = 1 -
               \left(\frac{r'_1}{r'_2}\right)^{N},
        \label{eq:07}
\end{equation}
where the second term corresponds to the term $\langle n \rangle P(0)$ 
term in Eq. (\ref{eq:2}).\\

\subsection{The Poisson transform}

The KNO scaling function $\Psi(z,t)$ can be also obtained by using the Poisson 
transform \cite{SB} in which it is related to the distribution
function $P(n,t)$ by the Poisson transform\\

\begin{picture}(400,30)(0,0)
       \put(280,10){$\Psi(z,t)$ .}
       \put(120,10){$P(n,t)$}
       \put(160,14){\vector(1,0){110}}
       \put(270,10){\vector(-1,0){110}}
       \put(180,18){\scriptsize  inverse  Poisson   trans.}
       \put(190,2){\scriptsize  Poisson   trans.}
 \end{picture}

\noindent
As a result we obtain in this approach that
\begin{equation} 
        P(n,t)=\int_{0}^{\infty}
               \frac{(\alpha \omega)^{n}}{n!}e^{-\alpha \omega}
               \Psi \left( \frac{\omega}{ \la n \ra /\alpha},t \right)
               \frac{d \omega}{ \la n \ra / \alpha} , 
\end{equation}
\begin{eqnarray}
     \Psi\left( \frac{ \omega}{ \la n \ra / \alpha},t \right)
           &=& \frac{1}{2 \pi}e^{\alpha \omega}
               \frac{\la n \ra }{\alpha}
               \int_{-\infty}^{\infty}
               e^{-ix \omega}
               \sum_{n=0}^{\infty} \left( \frac{ix}{\alpha} 
               \right)^n P(n,t)dx            
         \nonumber \\  
                &=& \frac{1}{2\pi i}\int_{\sigma-i\infty}^{\sigma+i\infty}
             e^{sz}
             \Pi(1-\frac{s}{\langle n \rangle},t)ds,
             \label{eq:09}
\end{eqnarray}
where $\Pi(u,t)$ is the generating function of $P(n,t)$.
These equations hold also in the stationary function($t=0$).

Using now the generating function Eq. (\ref{eq:0}), 
$\Psi(z)$ is given by the following inverse Laplace transform
\begin{eqnarray}
   \Psi \left( z\right)
         = \frac{1}{2\pi i}\int_{\sigma-i\infty}^{\sigma+i\infty}
             e^{sz}
             \left(\frac{r_1}{r_2}\right)^{N}
             \sum_{j=0}^{N}{}_{N}C_{j}
             \left( \frac{r_2-r_1}{r_1}\frac{\langle n \rangle}{r_2} \right)^{j}
             \left( s+\frac{\langle n \rangle}{r_2}\right)^{-j}ds.
                \label{eq:10}
\end{eqnarray}
Then we arrive at the KNO scaling function for the
MNBD\footnotemark[2] 
\footnotetext[2]{In the actual analysis the delta function term
occurring here will be approximated by the following expression
$$        \delta(z-\epsilon) 
             = \lim_{\alpha \rightarrow {\rm {large}}}
        \sqrt{\frac{\alpha}{2 \pi}}{\rm {exp}}[-\alpha (z-\epsilon)^{2}/2].     
$$
}
\begin{eqnarray}
  \Psi(z) &=& \left(\frac{r'_1}{r'_2}\right)^{N}\delta(z-\epsilon) +
             \left(\frac{r'_1}{r'_2}\right)^{N}
             e^{-\frac{\langle n \rangle}{r'_2}z}
             \sum_{j=1}^{N}{}_NC_{j} \frac{1}{\Gamma(j)}
             \left(\frac{r'_2-r'_1}{r'_1}\right)^{j}
              \left(\frac{\langle n \rangle}{r'_2} \right)^{j}z^{j-1}
          \nonumber \\
          &=& \left(\frac{r'_1}{r'_2}\right)^{N}\delta(z-\epsilon) +
              \left(\frac{r'_1}{r'_2}\right)^{N}
              \frac{r'_2-r'_1}{r'_2}\frac{\langle 
              n \rangle}{r'_1}
              e^{-\frac{\langle n \rangle}{r'_2}z}
              L^{(1)}_{N-1}
              \left(-\frac{r'_2-r'_1}{r'_2}\frac{\langle 
              n \rangle}{r'_1}z \right),
                \label{eq:08}
\end{eqnarray}
where 
$L_{n}^{(\alpha)}(x)$ is the Laguerre's polynomial.
In Eq. (\ref{eq:08}) the first term corresponds to the constant term 
in Eq. (\ref{eq:10})(, or $\langle n \rangle P(0)$).
Because the normalization is now just the unity,
in what follows we shall use this function as the KNO scaling 
function of the MNBD in the analysis of data.

\section{Analysis of experimental data}

We shall investigate now the applicability of the MNBD 
as presented in its KNO form (i.e., using Eq. (\ref{eq:08}))
to the description of the observed multiplicity distributions 
in $e^{+}e^{-}$ annihilations\cite{HRS}-\cite{opal96} 
and in hadronic collisions\cite{isr,ua587}.
Table I shows obtained parameters $C_2$, $N$ and the corresponding values of 
the $\chi^2_{\rm{min}}$
. 
Moreover it shows other parameters $r'_1$ and $r'_2$ calculated 
from $C_2$ and $N$ according to Eq. (\ref{eq:06}).

The corresponding KNO scaling functions for our MNBD are compared 
with the KNO form of the above mentioned multiplicity distributions 
in Fig. 1a - 1f and Fig. 2a - 2g.

\section{Summary and Discussion}

The KNO scaling function of the MNBD is obtained and is applied to the 
analysis of the observed multiplicity distribution in $e^{+}e^{-}$ annihilations 
and hadronic collisions.
The data are also analysed by the gamma distribution which is the KNO 
scaling function of the NBD.
As is seen from Table I, 
the  $\chi^{2}_{\rm{min}}$ values for the MNBD fit 
are almost equivalent to those for the NBD fit in both reactions.
The result shows to be similar to the case of the analysis of the cumulant 
moments in hadronic collisions\cite{NBSh}. 
On the other hand it should be noticed that 
the MNBD described the data of the cumulant moments 
much better than the NBD in $e^{+}e^{-}$ annihilations\cite{NBS}.
\\ \\

In order to know the stochastic structure of the MNBD in detail, 
we discuss the following point: 
The solution obtained
 from the branching equation of the pure birth process with the immigration
 under the initial condition of the binomial distribution\cite{NBSh,BSW} 
is one of the extensions of both the MNBD and the NBD. 
 Its generating function is given as
\begin{equation}
	\Pi(u) = \sum_{n=0}^{}P(n)u^{n}
	       = [1-r_1(u-1)]^{N}[1-r_2(u-1)]^{-k-N}.
	\label{eq:11}
\end{equation}
where $k$ is the immigration rate, and 
\begin{eqnarray}
   r_{1}&=&\frac{\langle n \rangle}{k}\left \{ 1 -\sqrt{\frac{k+N}{N}
            \left[ -k \left( C_2 -1-\frac{1}{\langle n \rangle} \right) 
            +1 \right] }~~\right \}
           \nonumber \\
    r_{2}&=& \frac{N r_1 + \langle n \rangle}{k+N}.\
\end{eqnarray} 
The MNBD is obtained by neglecting the power 
$k$ in $\Pi(u)$. The generating function of the NBD 
is given by neglecting the power $N$ in Eq. (\ref{eq:11}).
The physical meaning of the immigration term
$k$ may be interpreted as a possible contribution from gluons. 

Using Eq. (\ref{eq:09}), we have directly the KNO scaling function for Eq. 
(\ref{eq:11})
\\
\begin{eqnarray}
  \Psi(z) & = &  \left(\frac{r_1}{r_2}\right)^{N}
                 e^{-\frac{\langle n \rangle}{r_2}z}
                 \sum_{j=0}^{N}{}_NC_{j} \frac{1}{\Gamma(k+j)}
                 \left(\frac{r_2-r_1}{r_1}\right)^{j}
                 \left(\frac{\langle n \rangle}{r_2} \right)^{k+j}z^{k+j-1}
          \nonumber \\
          & = &  \frac{\Gamma(N+1)}{\Gamma(N+k)}
                 \left(\frac{r_1}{r_2}\right)^{N}
                 \left(\frac{\langle n \rangle}{r_2}\right)^{k}
                 z^{k-1}
                 e^{-\frac{\langle n \rangle}{r_2}z}
                 L^{(k-1)}_{N} \left( -\frac{r_2-r_1}{r_2}\frac{\langle 
                 n \rangle}{r_1}z\right).
           \label{eq:12}
\end{eqnarray}
This function becomes the KNO scaling function of the MNBD (Eq. 
(\ref{eq:08})) when $k \rightarrow 0$, and reduces to the gamma 
distribution (Eq. (\ref{eq:4})) if $N=0$.

In concrete application, we have confirmed that the discrete distribution 
of Eq. (\ref{eq:12}) can not explain
the oscillating behaviors of the cumulant moment observed
 in $e^+e^-$ annihilations and in hadronic collisions 
much better than the MNBD (Eq. (\ref{eq:2})).
However, we are expecting at present that Eq. (\ref{eq:12}) 
will become useful in analyses of data of some reactions at higher energies, 
since it has stochastic characteristics of the MNBD and the NBD.

\newpage
\vspace{5mm}
{\bf Acknowledgments}

One of the authors (M. B.) is partially supported by the Grant-in Aid 
for Scientific Research from the Ministry of Education, Science and 
Culture (No. 06640383) and (No. 08304024).  
N. S. thanks for the financial support by Matsusho Gakuen Junior College. 
We are grateful to G. Wilk for his reading the manuscript.

%
\newpage
\begin{flushleft}
{\large Table caption}
\end{flushleft}
\begin{itemize}
\item[Table I] 
The parameters of the KNO scaling function of the MNBD used in the 
analysis compared with those of the NBD for the charged 
multiplicity in: (a) $e^{+}e^{-}$ collisions and (b) hadronic collisions.
\end{itemize}
%
\vspace{1cm}
\begin{flushleft}
{\large Figure captions}
\end{flushleft}
\begin{itemize}
\item[Fig. 1]
The KNO form of the charged multiplicity in $e^{+}e^{-}$ collisions.
The full circles are obtained from data of the following 
collaborations:
HRS\cite{HRS}, TASSO\cite{tass89}, AMY\cite{amy}, 
DELPHI\cite{delp91} and OPAL\cite{opal92,opal96}, respectively, 
in Fig. 1a-1f.
The broken line is the KNO scaling function obtained from the 
MNBD.
\item[Fig. 2] 
The KNO form of the charged multiplicity in hadronic collisions.
The full circles are obtained from data of the following collaborations: 
ISR\cite{isr} and UA5\cite{ua587} collaborations, 
respectively, in Fig. 2a-g.
The broken line is the KNO scaling function obtained from the 
MNBD.
\end{itemize}
\newpage

\begin{center}
\begin{tabular}{l l r  
 r@{.}l @{ $\pm$ } r@{.}l
 r@{.}l @{ $\pm$ } r@{.}l
 r@{.}l @{ $\pm$ } r@{.}l
 r@{.}l @{ / }l}
        \hline
     Eq. (\ref{eq:08})
    &$\sqrt{s}~~{\rm[GeV]}$
    & N 
    &\multicolumn{4}{c}{$C_{2}$}
    &\multicolumn{4}{c}{$r'_{1}$}
    &\multicolumn{4}{c}{$r'_{2}$}
    &\multicolumn{3}{c}{$\chi^{2}_{{\rm min}}$/NDF}  \\
    \hline
  \hline
  HRS   & 29  & 13& 1&083 & 0&002 & 0&039 & 0&013 & 1&029 & 0&027 &32&5 &12 \\
  TASSO & 34.8& 11& 1&094 & 0&001 & 0&021 & 0&007 & 1&257 & 0&043 &39&0 &16 \\
  AMY   & 57  & 12& 1&084 & 0&003 & 0&006 & 0&026 & 1&438 & 0&048 &18&8 &18 \\
  DELPHI& 91.2& 11& 1&092 & 0&001 & 0&011 & 0&001 & 1&902 & 0&073 &79&0 &23 \\
  OPAL  & 91.2& 11& 1&092 & 0&001 & 0&012 & 0&015 & 1&957 & 0&042 & 4&5 &25 \\
        & 133 & 11& 1&093 & 0&004 & 0&029 & 0&001 & 2&156 & 0&060 & 7&2 &23 \\
        \hline
        \multicolumn{18}{c}{} \\
        \hline
     Eq. (\ref{eq:4})
    &$\sqrt{s}~~{\rm[GeV]}$
    &  
    &\multicolumn{4}{c}{}
    &\multicolumn{4}{c}{$k$}
    &\multicolumn{4}{c}{}
    &\multicolumn{3}{c}{$\chi^{2}_{{\rm min}}$/NDF} \\
    \hline
     \hline
  HRS   & 29  &&\multicolumn{4}{c}{}& 12&34 & 0&30&\multicolumn{4}{c}{}& 36&9 &13 \\
  TASSO & 34.8&&\multicolumn{4}{c}{}& 10&75 & 0&12&\multicolumn{4}{c}{}& 42&5 &17 \\
  AMY   & 57  &&\multicolumn{4}{c}{}& 11&90 & 0&43&\multicolumn{4}{c}{}& 19&1 &19 \\
  DELPHI& 91.2&&\multicolumn{4}{c}{}& 10&87 & 0&12&\multicolumn{4}{c}{}& 77&9 &24 \\
  OPAL  & 91.2&&\multicolumn{4}{c}{}& 10&86 & 0&17&\multicolumn{4}{c}{}&  4&4 &26 \\
        & 133 &&\multicolumn{4}{c}{}& 10&45 & 0&54&\multicolumn{4}{c}{}&  6&6 &24 \\
     \hline
\end{tabular}
%
\\
\vspace{1cm}
Table I (a)
\\
\end {center}
\vspace{2.5cm}
\newpage
\begin{center}
\begin{tabular}{l l r  
 r@{.}l @{ $\pm$ } r@{.}l
 r@{.}l @{ $\pm$ } r@{.}l
 r@{.}l @{ $\pm$ } r@{.}l
 r@{.}l @{ / }l}
        \hline
     Eq. (\ref{eq:08})
    &$\sqrt{s}~~{\rm[GeV]}$
    & N 
    &\multicolumn{4}{c}{$C_{2}$}
    &\multicolumn{4}{c}{$r'_{1}$}
    &\multicolumn{4}{c}{$r'_{2}$}
    &\multicolumn{3}{c}{$\chi^{2}_{{\rm min}}$/NDF}  \\
    \hline
    \hline
 ISR   & 30.4&  6& 1&190 & 0&015 & 0&123 & 0&079 & 1&880 & 0&083 &15&6 &15 \\
       & 44.5&  6& 1&198 & 0&005 & 0&189 & 0&030 & 2&203 & 0&038 & 5&2 &17 \\
       & 52.6&  9& 1&205 & 0&004 & 0&599 & 0&026 & 2&017 & 0&034 & 4&8 &19 \\
       & 62.2& 10& 1&195 & 0&004 & 0&647 & 0&028 & 2&010 & 0&036 &25&3 &18 \\
 UA5   & 200 &  4& 1&264 & 0&011 & 0&150 & 0&118 & 5&500 & 0&156 & 7&6 &29 \\
       & 546 &  4& 1&275 & 0&004 & 0&368 & 0&060 & 7&718 & 0&243 &54&2 &45 \\
       & 900 &  4& 1&301 & 0&009 & 0&908 & 0&024 & 9&808 & 0&257 &78&2 &52 \\
    \hline
        \multicolumn{14}{c}{} \\
        \hline
     Eq. (\ref{eq:5})
    &$\sqrt{s}~~{\rm[GeV]}$
    &  
    &\multicolumn{4}{c}{}
    &\multicolumn{4}{c}{$k$}
    &\multicolumn{4}{c}{}
    &\multicolumn{3}{c}{$\chi^{2}_{{\rm min}}$/NDF}
    \\
    \hline
        \hline
 ISR  & 30.4&&\multicolumn{4}{c}{}& 5&263 & 0&499&\multicolumn{4}{c}{}& 18&8 &16 \\
      & 44.5&&\multicolumn{4}{c}{}& 4&926 & 0&146&\multicolumn{4}{c}{}&  7&9 &18 \\
      & 52.6&&\multicolumn{4}{c}{}& 4&762 & 0&113&\multicolumn{4}{c}{}& 33&8 &20 \\
      & 62.2&&\multicolumn{4}{c}{}& 5&263 & 0&416&\multicolumn{4}{c}{}& 55&5 &19 \\
 UA5  & 200 &&\multicolumn{4}{c}{}& 3&788 & 0&158&\multicolumn{4}{c}{}&  7&6 &30 \\
      & 546 &&\multicolumn{4}{c}{}& 3&597 & 0&065&\multicolumn{4}{c}{}& 52&7 &46 \\
      & 900 &&\multicolumn{4}{c}{}& 3&257 & 0&095&\multicolumn{4}{c}{}& 62&8 &53 \\
        \hline
\end{tabular}
\\
\vspace{1cm}
Table I (b)
\end{center}
\newpage
\begin{figure}
\psbox[scale=0.8]{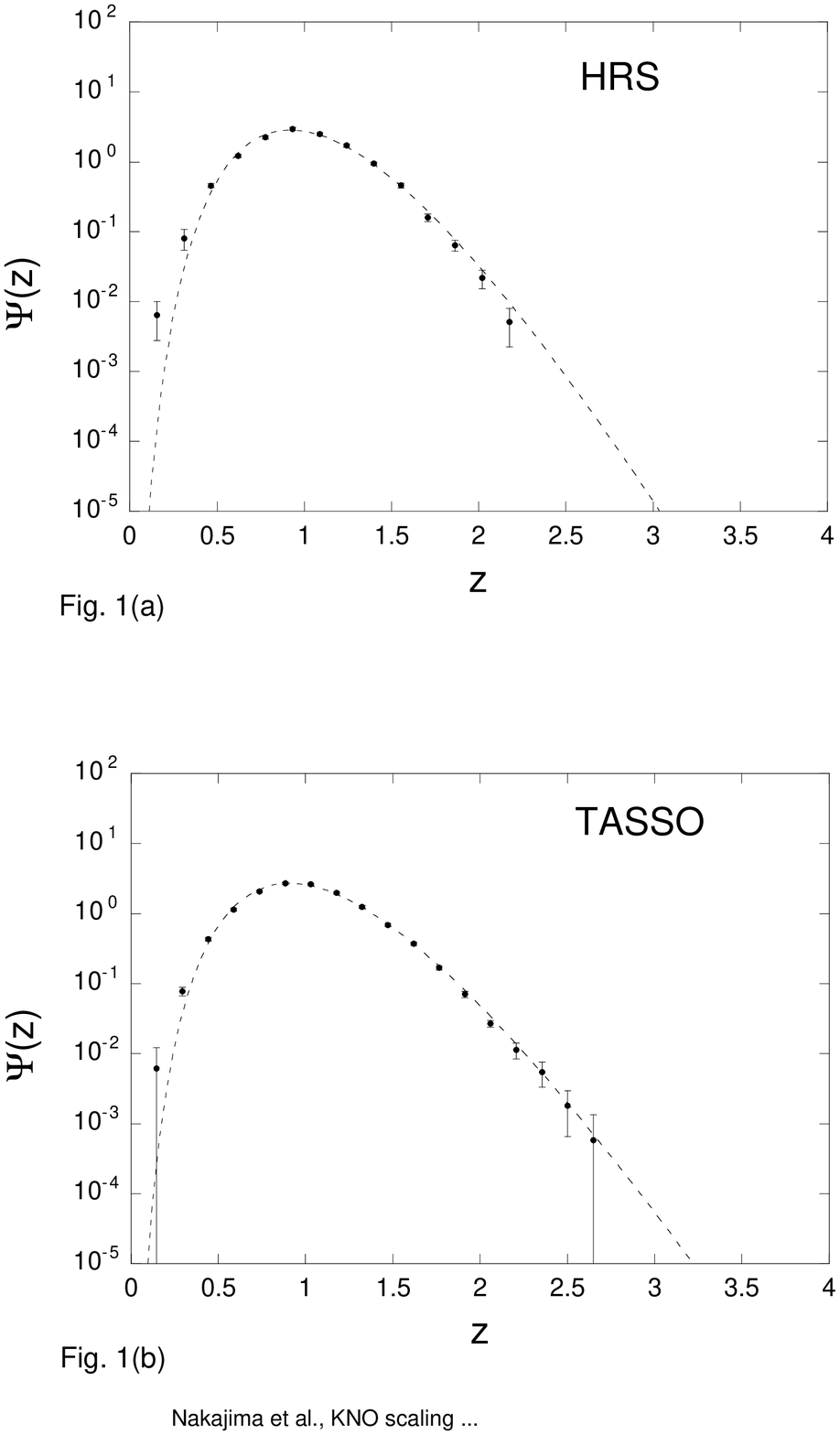}
\end{figure}

\newpage
\begin{figure}
\psbox[scale=0.8]{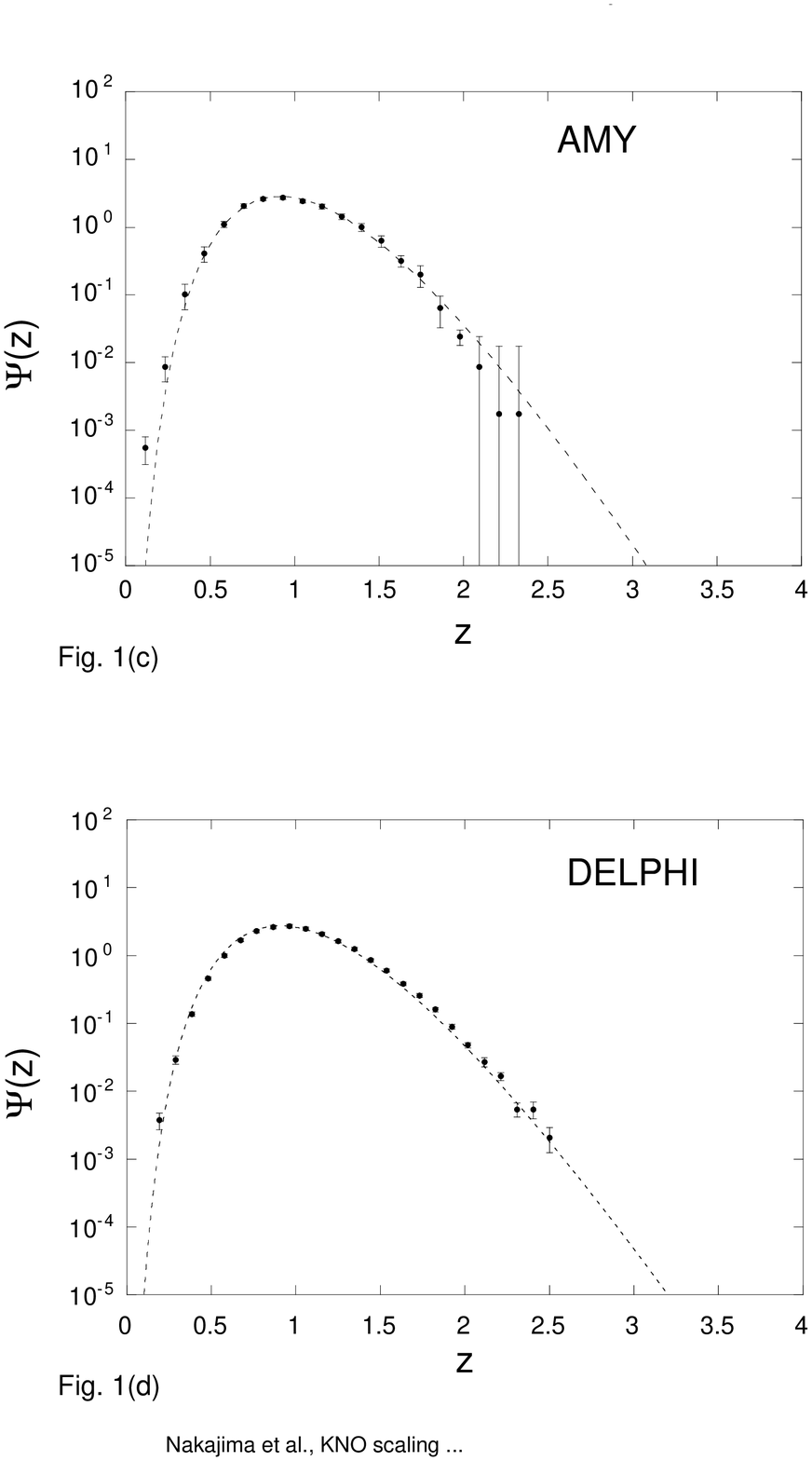}
\end{figure}

\newpage
\begin{figure}
\psbox[scale=0.8]{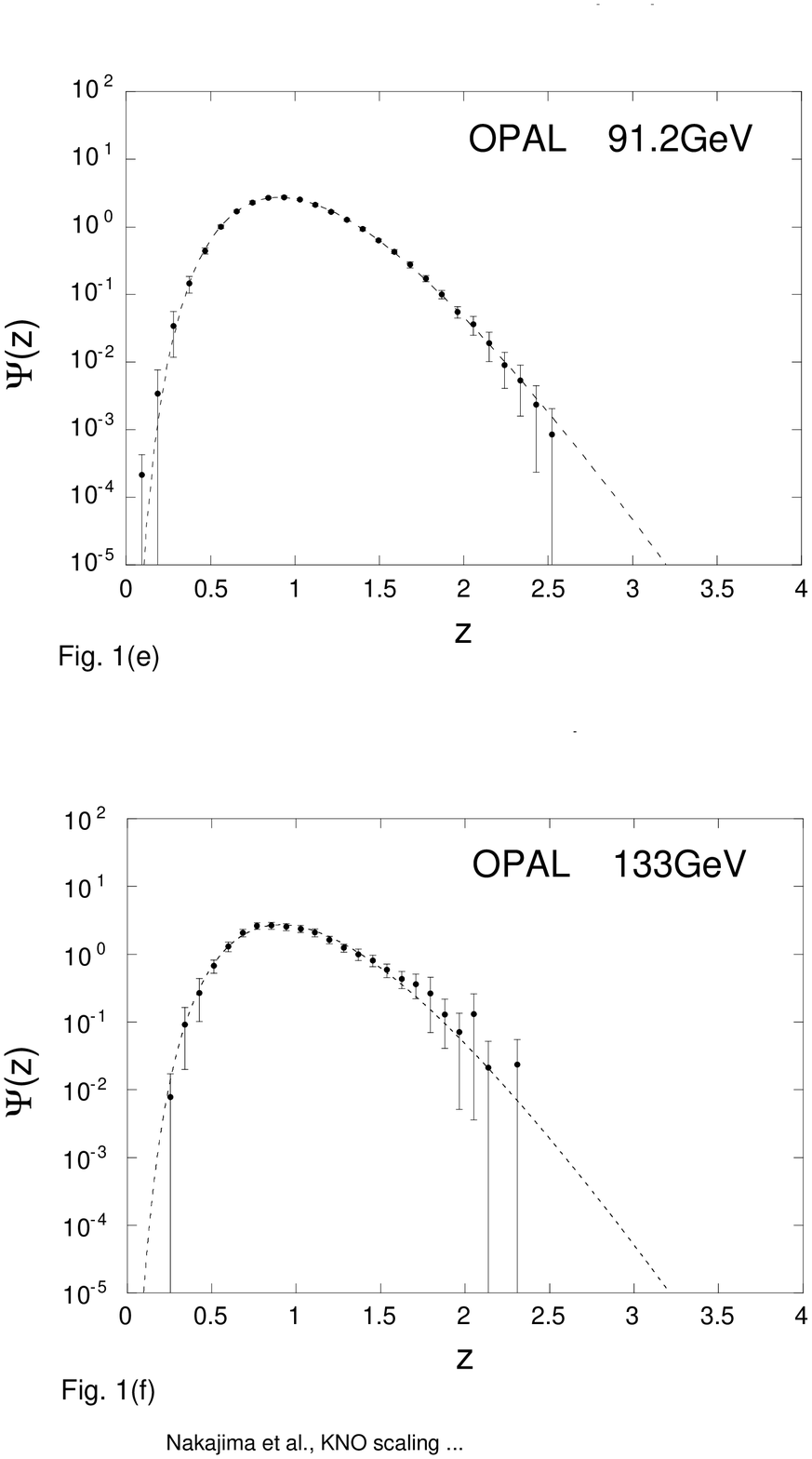}
\end{figure}

\newpage
\begin{figure}
\psbox[scale=0.8]{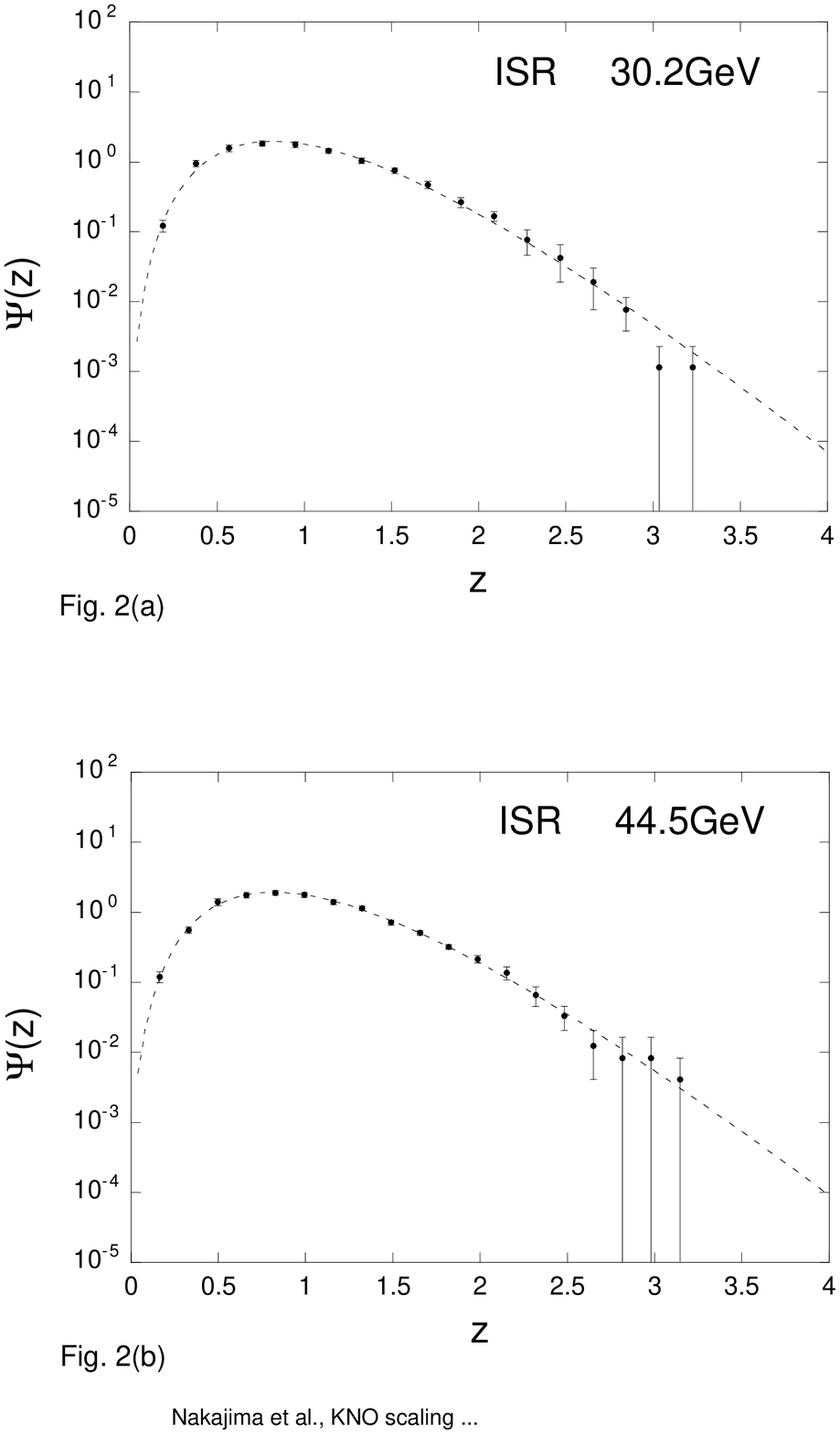}
\end{figure}

\newpage
\begin{figure}
\psbox[scale=0.8]{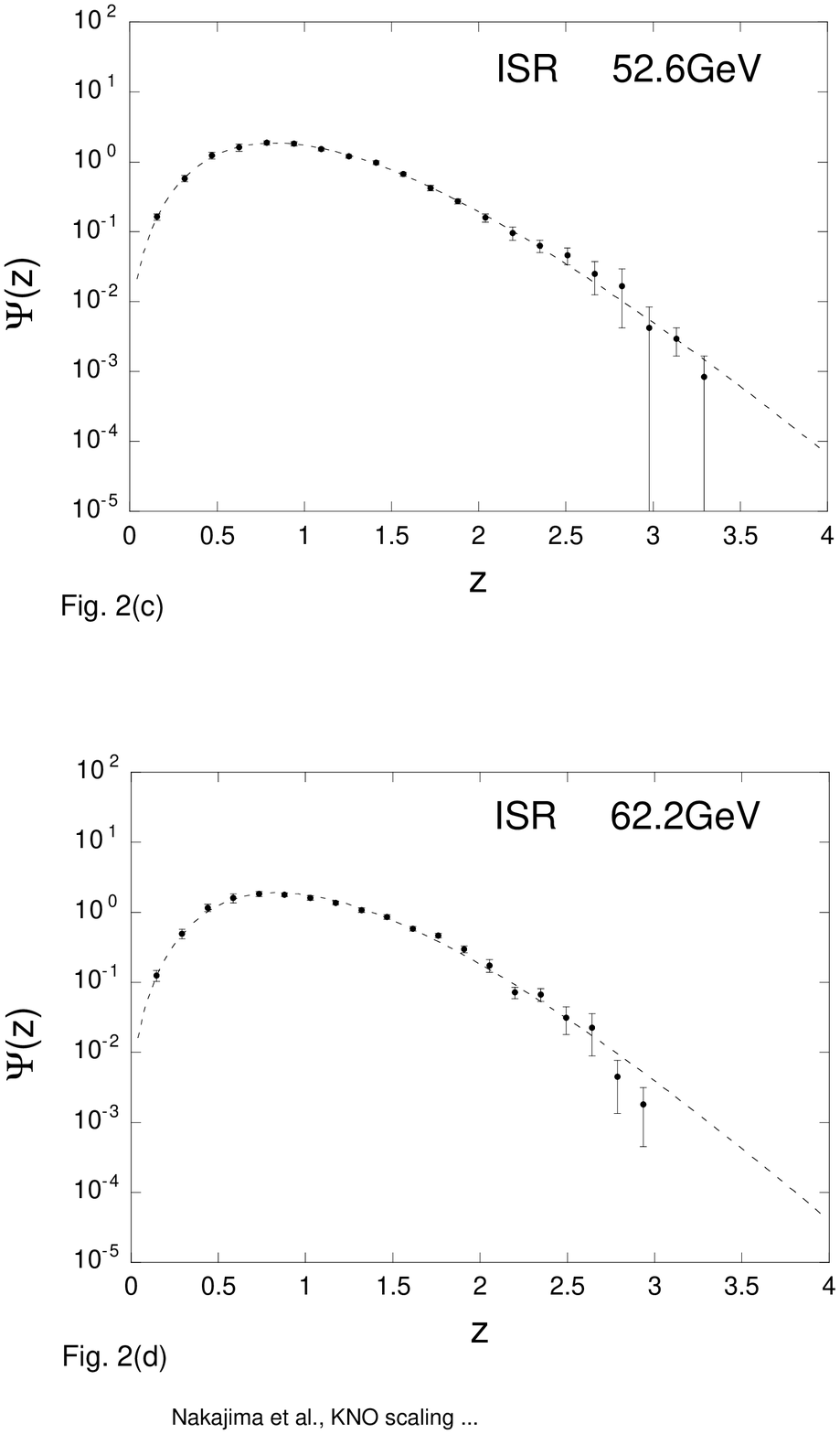}
\end{figure}

\newpage
\begin{figure}
\psbox[scale=0.8]{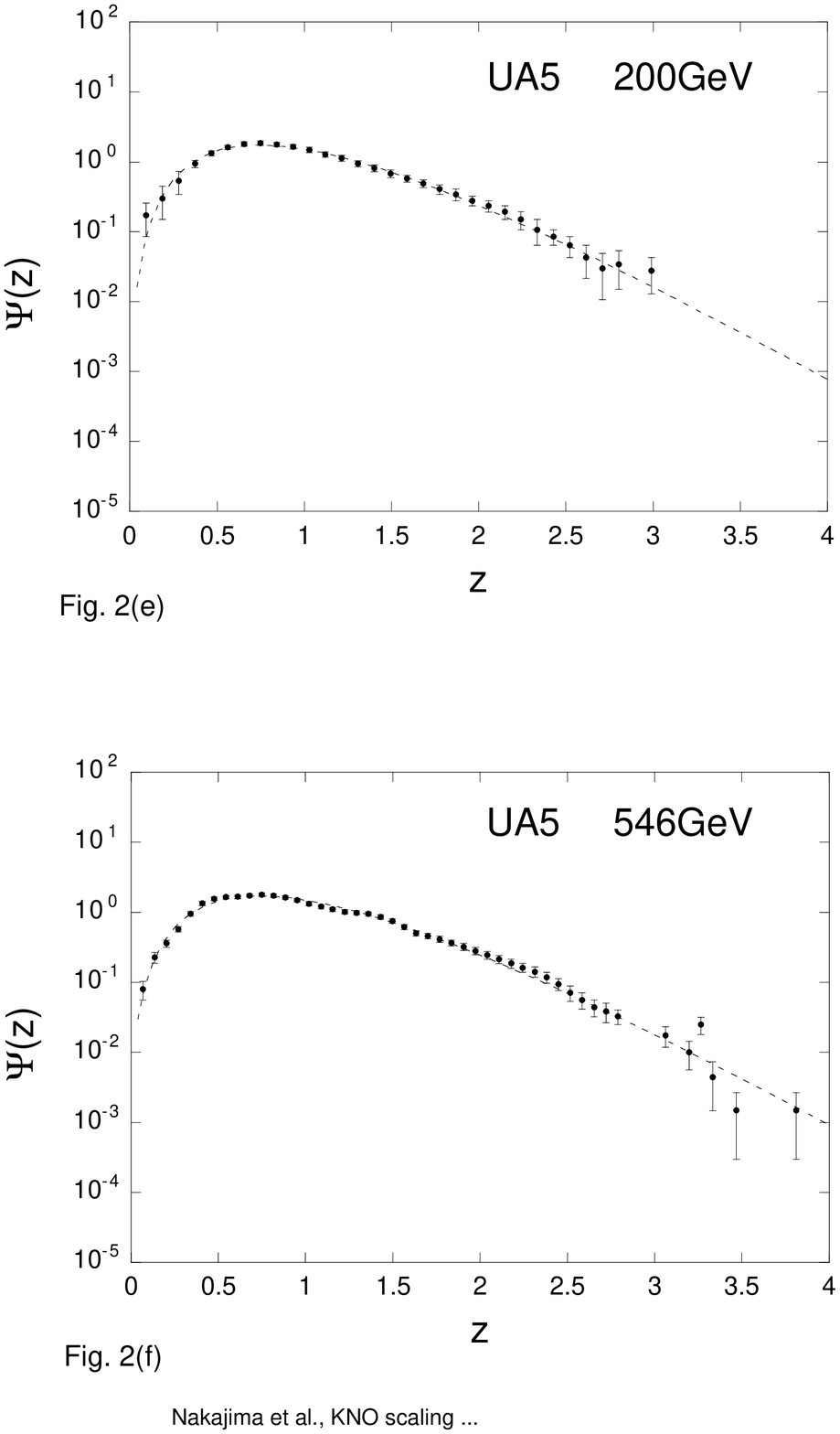}
\end{figure}

\newpage
\begin{figure}
\psbox[scale=0.8]{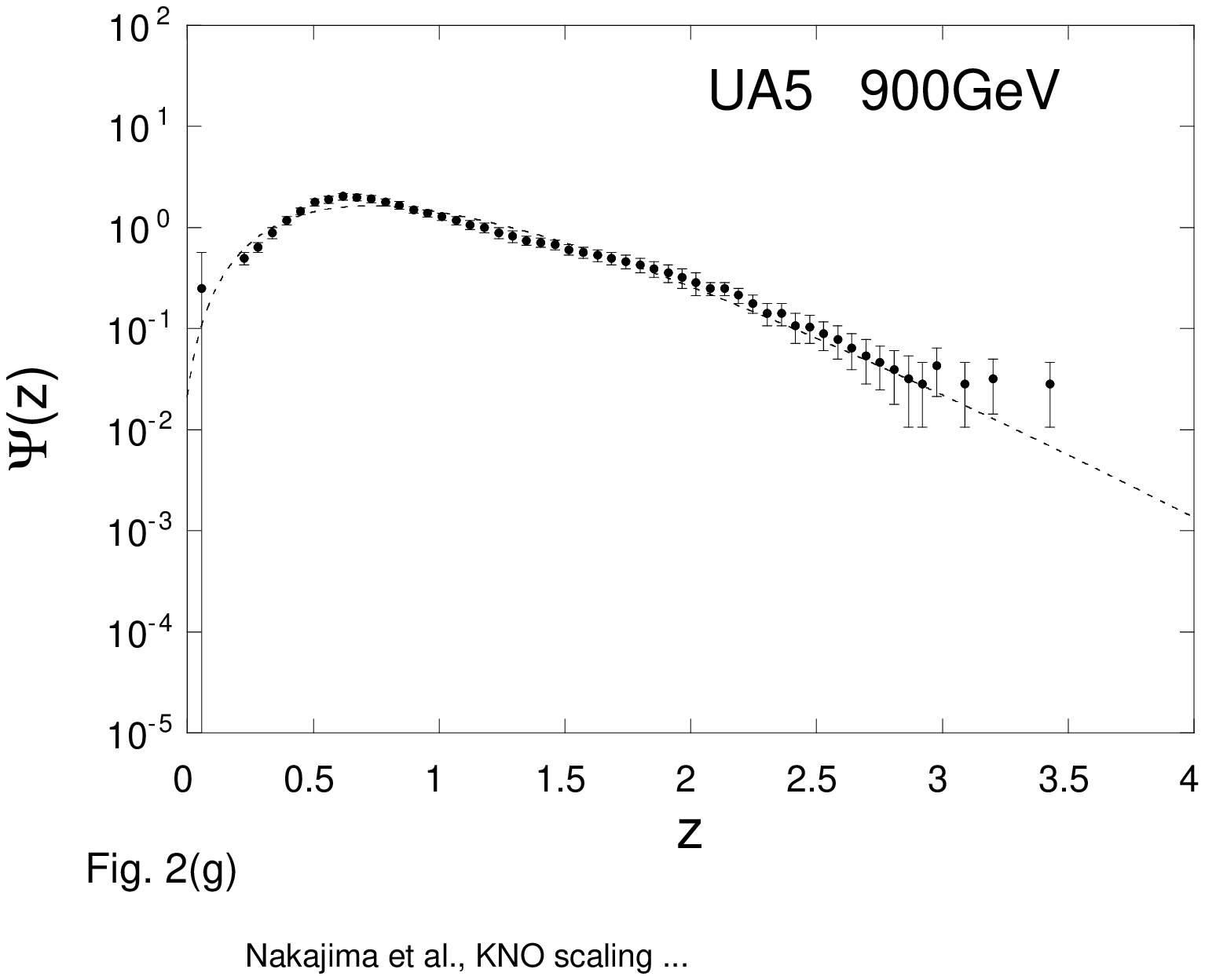}
\end{figure}

\end{document}